\documentclass[10pt,journal,compsoc]{IEEEtran}
\usepackage{rotating}
\usepackage{graphicx}
\usepackage[table,xcdraw]{xcolor}
\usepackage{tikz}
\usepackage{capt-of}
\usepackage{algorithm}
\usepackage[utf8]{inputenc}
\usepackage[noend]{algpseudocode}
\usepackage{epsfig,array}
\usepackage{amsmath,amssymb}
\usepackage{amsfonts}
\usepackage{graphicx}
\usepackage{color}
\usepackage{epstopdf}
\usepackage{mathtools}
\usepackage{float}
\usepackage{subfigure}
\usepackage{placeins}
\usepackage{balance}
\usepackage{hyperref}
\usepackage{accents}
\usepackage{adjustbox}
\usepackage[margin=2cm]{geometry}
\usepackage{rotating}
\usepackage{ulem}
\usepackage{dblfloatfix}

\usepackage{array}
\newcolumntype{P}[1]{>{\centering\arraybackslash}p{#1}}
\usepackage{xspace}

\usepackage{comment}

\usepackage{textcomp}
\usepackage{xcolor}

\ifCLASSINFOpdf
 
\else

\fi

\begin{document}

\title{Decentralized Zero-Trust Framework for Digital Twin-based 6G}

\author{\textbf{Ismaeel Al Ridhawi},~\IEEEmembership{Senior Member,~IEEE,}
        \textbf{Safa Otoum},~\IEEEmembership{Member,~IEEE,}
        \textbf{Moayad Aloqaily},~\IEEEmembership{Senior Member,~IEEE}
\thanks{I. Al Ridhawi is with the University of Ottawa, Canada. \protect E-mail: ismaeel.alridhawi@uottawa.ca}
\thanks{S. Otoum is with Zayed University, UAE. \protect E-mail: safa.otoum@zu.ac.ae}
\thanks{M. Aloqaily is with MBZUAI, UAE. E-mail: \protect moayad.aloqaily@mbzuai.ac.ae}

}

\maketitle
\begin{abstract}
The Sixth Generation (6G) network is a platform for the fusion of the physical and virtual worlds. It will integrate processing, communication, intelligence, sensing, and storage of things. All devices and their virtual counterparts will become part of the service-provisioning process. In essence, 6G is a purposefully cooperative network that heavily depends on the capabilities of edge and end-devices. Digital Twin (DT) will become an essential part of 6G, not only in terms of providing a virtual representation of the physical elements and their dynamics and functionalities but rather DT will become a catalyst in the realization of the cooperative 6G environment. DT will play a main role in realizing the full potential of the 6G network by utilizing the collected data at the cyber twin and then implementing using the physical twin to ensure optimal levels of accuracy and efficiency. With that said, such a cooperative non-conventional network infrastructure cannot rely on conventional centralized intrusion detection and prevention systems. Zero-trust is a new security framework that aims at protecting distributed data, devices, components and users. This article presents a new framework that integrates the zero-trust architecture in DT-enabled 6G networks. Unlike conventional zero-trust solutions, the proposed framework adapts a decentralized mechanism to ensure the security, privacy and authenticity of both the physical devices and their DT counterparts. Blockchain plays an integral part in the authentication of DTs and the communicated data. Artificial Intelligence (AI) is integrated into all cooperating nodes using meta, generalized and federated learning solutions. The article also discusses current solutions and future outlooks, with challenges and some technology enablers. 
\end{abstract}

\begin{IEEEkeywords}
AI, DT, 6G, Zero-Trust, Moving Trust, General Trust.
\end{IEEEkeywords}

\IEEEpeerreviewmaketitle

\section{Introduction}
\IEEEPARstart{W}{ith} device and technology transformation, our reliance on Information Technology (IT) solutions, devices, systems, networks, and processes grow increasingly. The Sixth Generation (6G) network is being introduced as purposeful network, where the network is derived from the use-cases, rather than the other way around. Such a network will enable massive automation of society and sustainable development. 6G applications and services will no longer be bound by the limitations of the network performance, size, cost, and power, as with its predecessors. Rather, 6G will provide full dimensional coverage which enables multi-access continuity. This is only achievable through a cooperative, distributed, trustworthy and intelligent global ecosystem. 

Digital Twin (DT) has become an integral part of Next-Generation Networks (NGN), especially for 6G. DT will play a significant role in the enablement of real-time monitoring, analysis, testing, and management of assets, especially those that are hard to reach. DT provides a replica of physical assets, processes, people, systems and devices. As such, it provides both the elements and dynamics of IoT device operation and is synchronized adaptively throughout the life cycle of the device. DT will create a metaverse of people and things through significant real-time data collection, analysis, and processing. This may not be possible without full dimensional coverage that enables for seamless multi-access continuity, as in 6G. Such DT-enabled 6G will support real-time scanning of the physical environment and creating a Three-Dimensional (3D) virtual environment (\textit{i.e.,} DT-supported metaverse) on-the-go.

\begin{figure*}[ht]
    \centering
    \includegraphics[scale=.4]{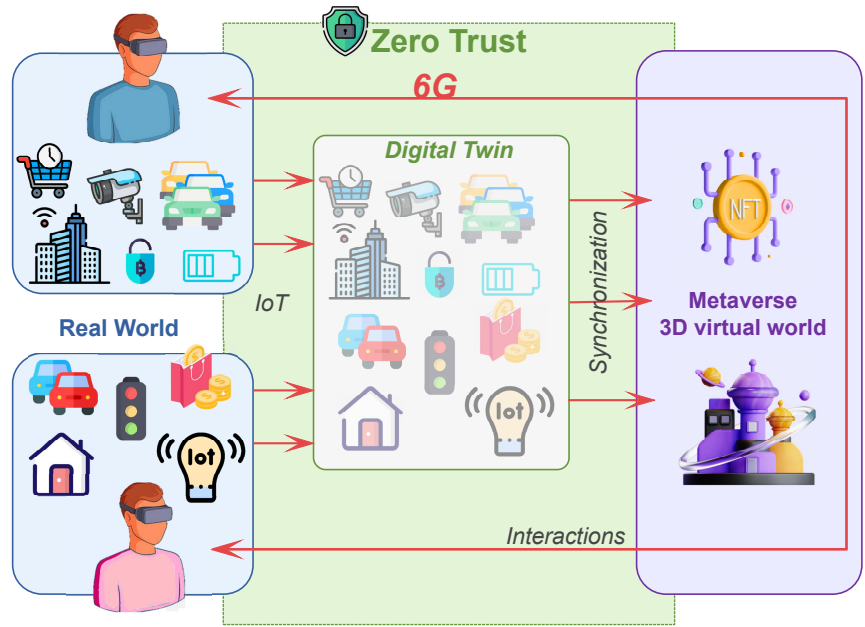}
    \caption{An Overview of the Zero-Trust Architecture in a 6G network environment.}
    \label{fig:fig01}
\end{figure*}

With DT-enabled 6G networks, enormous numbers of devices will be involved in the communication, data sharing, and service provisioning process. In such a network environment, trust and security cannot be maintained nor managed through conventional Intrusion Detection Systems (IDS) and Intrusion Prevention Systems (IPS). The Zero Trust architecture has been introduced lately as an alternative solution to conventional IDS and IPS. In zero trust, it is assumed that non of the elements nor the key players in a system are to be trusted, as depicted in Figure \ref{fig:fig01}. Unlike conventional solutions, where a single network approach is considered, such that devices in the network are assumed to be trustworthy, zero trust architectures are considered to be not a single network architecture. A composite of various security principles and strategies are adopted to ensure distributed data, devices, components and users are secure and authentic. Zero-Trust Architectures do not rely on entry-point authentication, as seen in firewalled networks. But rather, the use of context-aware, dynamic, and intelligent authentication schemes is a must to detect and prevent cyber-security intrusions and attacks. The high-level overview of the zero trust architecture depicted in Figure \ref{fig:fig01} highlights the need for zero trust frameworks that not only secure 6G devices, but rather the DTs of the physical devices, the virtual interactions between the DTs and the communication infrastructure used to enable metaverse interactions of DTs. 

Two crucial elements should be part of novel Zero-Touch architectures, namely, threat intelligence and decentralized authentication. Threat intelligence serves as an autonomous security strategy to ensure continuous trust evaluation and attain high-levels of access control within the networked environment. Artificial Intelligence (AI) will support DT-enabled 6G networks to maintain network trustworthiness in terms of security, safety, resilience, privacy, and reliability on the move for both the network environment and operations. Moreover, decentralized authentication is becoming more feasible as we near the 6G era through solutions such as Blockchain. Additionally, blockchain can avoid data leakage when exchanging data and obscure malicious operations 6G's massive device interconnection. Blockchain will play a significant role in the decentralized zero-trust architecture and can provide a solution to network scalability in 6G while maintaining decentralized network management.

\begin{figure*}
    \centering
    \includegraphics[scale=.5]{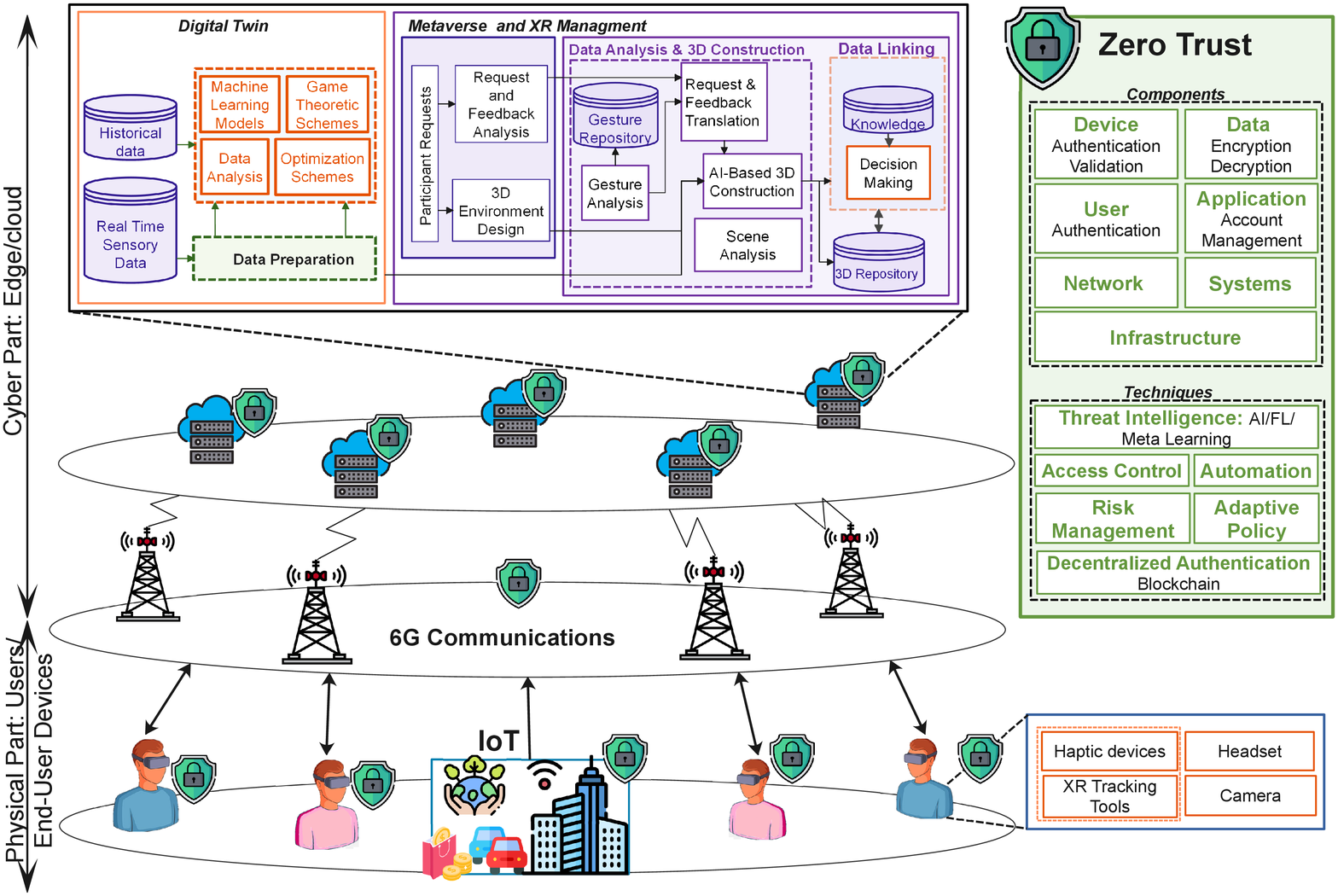}
    \caption{A proposed architecture}
    \label{fig:fig02}
\end{figure*}

Figure \ref{fig:fig02} highlights some of the zero trust architecture components and techniques used to maintain high levels of zero trust management. The zero trust architecture is decentralized and is adopted at the IoT device level, the 6G communication level, as well as the DT of the users, devices, and data. In essence, the framework must ensure device authentication and validation, data encryption and decryption, user authentication, application account management, network, system, and infrastructure security. Such a framework will require decentralized access control, high levels of automation, self-configuring adaptive authentication policies, risk management, and most importantly, threat intelligence and decentralized authentication. 

\section{Zero-Trust Through Threat Intelligence}
\subsection{Current Trends}
In a 6G environment, especially with one that is supported by DT, with the massive number of participating devices, whether real or DT, it becomes nearly impossible to use common endpoint security controls to secure the network environment. Moreover, in 6G the network will become more prone to various types of cyber-security threats and attacks \cite{ref1}. To detect and prevent cyber attacks, intrusion detection methods that rely on the use of Machine Learning (ML) and Deep Learning (DL) have shown significant effectiveness in terms of threat detection rate \cite{ref2}. Such solutions rely on data collection and labeling, which is somewhat complex in terms of time and cost. In a 6G environment, it will become nearly impossible to adopt such an approach with rapid changes in cyber-attacks and the explosive number of IoT devices. Furthermore, relying on centralized ML and DL solutions is not sustainable from a communication and processing perspective. Edge and end-devices cannot continuously share the collected devices and wait for predictive results in time-critical scenarios.

In \cite{ref2}, the authors developed a semi-supervised Deep Reinforcement Learning (DRL) algorithm to detect abnormal network traffic. Two unsupervised learning algorithms, namely, Auto Encoder and K-means are combined to reconstruct network traffic features. Instead of relying on a real network environment, the solution uses a simulated environment instead to apply the DRL. Such an approach makes it simple and adaptable for high-demand network environments. Fast training and prediction are achievable in real-time to detect and prevent cyber-attacks. Another DL solution was developed in \cite{ref3} to detect cyber threats in Space-Air-Ground Integrated Networks (SAGIN). Hidden traffic patterns of malicious events are extracted using a deep sparse autoencoder algorithm. Then, a Gated Recurrent neural network (GRNN) is used to identify abnormal behaviours and identify malicious attack types. 

\subsection{Potential Solutions}
Distributed learning solutions that rely more on edge and end-devices can play a significant role in securing 6G networks. Meta-learning is a new approach to learning that adaptively changes the ML algorithms run on devices to improve the model performance and accuracy. Simply put, it is the concept of learning to learn. In essence, given that different learning algorithms and datasets may lead to different result accuracies, meta-learning learns to apply the most accurate results from different ML algorithms, metadata and outputs to improve its model for future runs. Given that 6G devices will be susceptible to a plethora of types of cyber attacks, an edge DL approach may not be able to handle all types of attacks. This is due to \textit{i)} not having sufficient dataset to perform the necessary training and evaluation, \textit{ii)} not able to learn new tasks with the limited set of attacks known at the edge device, and \textit{iii)} the complexity in terms of computation and time for running a predictive algorithm at a single node while serving a significant number of end-devices.

General models applied at different edge sites and then later scaled to accommodate for new attack types based on a meta-learning approach can resolve conventional learning issues. Such an approach will allow for training to tackle multiple models using multiple datasets, in essence, learning may be achieved with fewer data at a much faster pace. More predictive classes may be added without needing to change the model structure. In \cite{ref4}, a meta-learning algorithm is used to detect WiFi impersonation attacks. The algorithm learns to adapt to new attack scenarios with the least dataset samples required. Multiple auxiliary networks are built to transfer meta-data for new attack detection tasks. The approach is highly favourable when compared to Deep Neural Network (DNN) algorithms, which require continuous time-consuming training from large dataset samples. DNN is not adaptable in real-time, whereas the proposed meta-learning algorithm can achieve accuracies of up to 98\% with less than 0.1\% of the training samples.

Federated Learning (FL) was introduced to overcome centralized learning shortcomings such as the need for significantly large datasets, processing and storage resource requirements, and most importantly the need to communicate collected data, which put user privacy and data security at risk. FL distributes the learning task between different edge and end participant nodes. Those nodes will not communicate the collected data, but rather perform localized model training on the local dataset. Then, it shares only the updated model parameters with the a central node. The central node then creates a global model using the shared local models. The global model parameters are then shared once again back to the participating nodes in the FL process. In \cite{ref5}, the authors developed a federated deep learning algorithm for intrusion detection in Cyber-Physical System (CPS). The approach enables an intrusion detection model to be built with the support of multiple industrial CPSs while preserving data privacy. A crypto-inspired communication protocol was also developed to maintain high levels of security and privacy when communicating the learnt parameters. 

Although both meta-learning and FL are promising solutions for use in 6G networks, the learning approaches are not yet mature enough to maintain trust and privacy in an environment that not only should secure physical devices and data, but also virtual devices and data in the form of DT. 6G networks will rely on distributed and decentralized infrastructures and management. In essence, with the vast number of devices and their DTs, decentralized zero trust solutions that rely on decentralized or layered distributed learning may provide an alternative to current distributed solutions that highly rely on centralized management. 

\subsection{Future Directions}
Decentralized zero trust may only be accomplished through high-levels of ML distribution and generality. Although meta-learning provides some sort of generality to the learning models, the solution cannot be applied to IoT end-devices. The model is assumed to move from general to high levels of complexity as new attacks types are added to the training model. In such a scenario, most participating IoT devices will not be capable of applying a meta-learning solution given that their resources might be constraint. Furthermore, participating devices in the meta-learning process are assumed to be managed globally so that metadata and outputs are shared in order to improve device local model for future runs. This introduce new data security and privacy issues that may lead to the failure of the zero trust architecture. With that said, decentralized zero trust can be maintained using ML generality which is adopted on most 6G IoT devices and through hierarchical FL approaches. 

In \cite{ref6}, the authors presented a generalized-AI solution that that is adaptable to IoT devices in all their forms. Meaning, that resource-limited IoT device such as sensors and actuators, as well as more resource-rich devices such as user end-devices and edge devices. These devices will become part of the service-provisioning process. Federated learning and meta-learning is adaptable at a scalable level only if all devices are participating in the learning process. Moreover, the learning may be adapted at the DTs as well. With the said, the authors showed that learnability is adapted at two levels of the system, namely, at the management and problem domain layers. Datasets and models are selected based on the input data and problem domain. With that said, this learning solution needs to accurately identify the configurations and learning-related tasks for the given problem. Then, at the problem domain layer, the adequate ML algorithms are selected in accordance to the identified problem and preferences. The authors argue that such a solution can only be achieved if the training data is distributed over all network participants, where such data is collected and preserved locally. Moreover, they indicate that the trained models must be collaboratively shared in the network for faster convergence. Edge devices may support IoT devices not only in the learning process but also in the data abstraction, filtering and reduction process. In essence, if those elements are made possible a plug-and-play AI (PnP-AI) solution may be realized. In essence, this could provide significant support to realize a decentralized zero-trust architecture.

In \cite{ref7}, the authors presented a Hierarchical Federated Learning (HFL) algorithm that is adaptable to CPS. The solution relies on FL, blockchain, and a trustworthy cooperative technique to establish a distributed learning environment for CPS management. Given that in a zero trust architecture all devices are assumed to be non-trustworthy, a decentralized trust scoring solution must be developed. With that the authors have defined device trustworthiness using a fuzzification algorithm that considers \textit{i)} cooperativeness level of the device in terms of task load, available resources and complexity of the tasks, and \textit{ii)} device characteristics in terms of security encryption capabilities, mobility, and configurations. The FL occurs at different layers and can be managed in a distributed or decentralized manner. IoT devices can perform learning locally and then FL is applied at the cluster level between grouped devices. FL is also applied at the fog level based on the shared cluster models which have been securely added to a blockchain. The framework also enables offline centralized learning to run on the cloud. Such a solution provides the possibility for zero trust to be managed in a decentralized manner.

\section{Zero-Trust Through Intelligent Blockchain}
In 6G networks, $\approx$6 billion connected mobile phone devices will be using the enhanced Mobile Broadband (eMBB) service that require latency not exceeding 10$ms$. As for the Ultra-Reliable Low Latency (URLCC) communication, $\approx$2 billion connected devices will require latencies not exceeding 1$ms$ for autonomous driving and tasks achievement services. Finally, $\approx$ 1 trillion IoT devices will use the massive-Machine-Type Communication (mMTX) service which will require latencies between 10$ms$ and 10$s$. From those devices $\approx$2 billion devices will have AI capabilities. Such an enormous number of devices cannot rely on centralized conventional security and zero trust systems. Moreover, a significant number of devices, their functionalities and processes, as well as the communicated data will whether in the physical world or on the metaverse will require rejuvenated time-critical and scalable trustworthy and secure communication.

\subsection{Current Trends}
With that said, the objective in DT-enabled 6G communication is to ensure trustworthy connectivity between the devices, whether physical or DT both in the physical world and on the metaverse. To do so, decentralized or highly-distributed and hierarchical solutions be considered. Blockchain is a promising solution that can obscure malicious operations in 6G's massive device interconnection. Blockchain technology was initially presented as decentralized solution to financial transactions but has lately been a hot research topic for different applications such as electronic healthcare and industry 4.0 \cite{ref8}. The use of blockchain in maintaining secure communication in NGNs has been studied for sometime, especially in Ad Hoc networks such as Vehicular Ad Hoc Networks (VANET) and Flying Ad Hoc Networks (FANET) \cite{ref9}.

In \cite{ref10}, the authors introduce a blockchain-based solution to authenticate 5G-enabled drones that roam through different network domains. The solution maintains information security and privacy for drone communication. 5G networks would not be able maintain data and device integrity for drones when crossing over multiple network domains. As such, the proposed solution create local private blockchains for local domain drone registration, authentication, and audit. Only drones that are authorized on the network will have access to communication and data through data blocks. Moreover, multi-signature smart contracts are used to maintain authentication across different network domains using consortium blockchain. In essence, the solution tackles storage limitations of blockchains and authentication latency in such multiple cross domain blockchain-enabled 5G network. 

In \cite{ref11}, the authors adopt using blockchain to secure 5G-enabled industrial IoT. The authors point out that although cloud-based controlling and process monitoring is highly beneficial, cloud integration will introduce security risk and cyber threats that will make the entire manufacturing system exposed. As such they opt to using fog-based solutions to eradicate message-routing threats. Blockchain maintains trust establishment through dynamic certificate creation to register cloud, fog, and IoT nodes. Moreover, symmetric keys are established between network nodes to encrypt messages and placed them on the blockchain. A zero-knowlege proof-based verification mechanism is used to maintain anonymity and unlinkability. To resolve blockchain storage issues, extended storage is integrated into the blockchain.

\subsection{Potential Solutions}
One of the main issues of blockchain is its lack of effectiveness in scalable network environments. Adding information to the blockchain is not the issue, but rather the consensus process is extremely time-sensitive as the size of the blocks and blockchain increase. A number of promising solutions have been introduced to make blockchains more intelligent. Such solutions may incorporate blockchains at the infrastructure, functional, and service operation levels. Also, categorizing services to create different consensus groups. Voters may belong to certain groups and participate in the consensus process. This will accelerate the voting process and enable blockchain to be adapted for scalable networks.

In \cite{ref12}, the authors presented a block transport solution to be adapted to scalable hyperledger fabric blockchain. The blockchain is used for device-to-device assisted 5G networks. The solution distributes the block data to all blockchain peers. The core functionalities of the blockchain are distributed among edge servers and mobile devices as well. As such, the servers will handle block generation and membership management. The overlay structure of the blockchain is organized by the edge servers. Mobile devices are responsible for endorsement and validation. Such a solution can significantly increase transaction throughput paving the road for blockchain to be used in delay-sensitive 5G services and applications.

In \cite{ref13}, the authors propose a purely decentralized blockchain consensus algorithm based on the Nakamoto protocol that achieves high levels of security and scalability. The algorithm generates two types of blocks, namely, micro- and macro-blocks. Micro-blocks are dedicated for in network transaction serialization, while macro-blocks are for chain formation. Using Proof of Work (PoW), a leader is elected during each round. Mined micro-blocks are placed into one macro-block by different nodes. This way, nodes are encouraged to participate in the transaction serialization process while limiting the leader's role. In essence, the solution will on miners rather than leaders, hence, producing a decentralized blockchain network. To further limit the leader's role, a transaction diversity-based metric is used for choosing the branch legality. Such a solution provides enhanced scalability while preserving the decentralization nature of blockchain.

\subsection{Future Directions}
A key technique used to achieve decentralized zero trust is to maintain an intelligent purely decentralized blockchain solution that can be adapted to scalable networks. One of the main pillars of 6g is being AI-native. Such characteristic cannot be attained without involving all devices and DTs in the learning process. As such, a decentralized zero trust solution can only be adapted by relying on the intelligence nature of 6G devices. Intelligent blockchain-enabled communication is a primary solution towards securing decentralized infrastructures. It is highly anticipated that 6G networks will require the use of blockchain not only to secure data and authenticate devices, but also become part of the authentication process of every element in 6G, whether that is a device, DT, process, function, or infrastructure.

\begin{table*}[h!]
    \centering
        \caption{A comparison between the different features of traditional and Zero-trust security schemes}
    \begin{tabular}{|p{1.4cm}||p{4cm}|p{4cm}|p{4cm}|}
    \hline
    \hline
         & \textbf{Traditional Security}  &\textbf{Centralized Zero-Trust}t& \textbf{Decentralized Zero-Trust} \\
         \hline
       \textbf{Device Trust}  & No frequent device validation & Frequent device validation& Frequent device validation\\
       &Centralized Verification & Centralized Verification& De-centralized Verification\\       
         \hline
    \textbf{Cryptography} & Centralized encryption & Highly controlled environment & End-to-end encryption\\
     & Full-disk encryption (FDE)&Centralized key management&Distributed key management\\
         \hline
       \multicolumn{1}{|p{1cm}||}{\textbf{Access Control}}  & Everything within the organization’s trusted &\multicolumn{2}{p{8cm}|}{Granting the least access needed to accomplish a specific task}\\
       &  &\multicolumn{2}{p{8cm}|}{High secure access to resources}\\
       \cline{3-4}
       &No further verification's required&Managed by a central entity&Distributed management\\
         \hline    
    \textbf{Authentication}  & Once at the beginning of the service request & Frequent during the service deliver & Very frequent during the service delivery\\
    &Password& Managed by a central entity &Distributed Auto-authenticate\\
    &PIN (Personal Identification Number)&Enhanced secured access&Least privilege access principles and Push notification authentication\\  
    \cline{3-4}
    &&\multicolumn{2}{p{8cm}|}{Single Sign- On (SSO)/Multi-factor authentication}\\
    \hline   
    \multicolumn{1}{|p{2cm}||}{\textbf{Threat Detection}}& Malicious activity identification &\multicolumn{2}{p{8cm}|}{Continuous monitoring, auditing/Continuous user behavior analytics}\\
     & Post-detection capabilities&\multicolumn{2}{p{8cm}|}{Proactive detection of malicious attacks}\\
         \hline
    \multicolumn{1}{|p{2cm}||}{\textbf{Risk Management}}  & Reduce risk exposure by avoiding known risks &\multicolumn{2}{p{8cm}|}{Continuous monitoring, assessments, and identification of components risk }\\
    &Counteract measurement&\multicolumn{2}{p{8cm}|}{Advanced risk management}\\
         \hline     
    \textbf{Automation} & Protecting the perimeter using firewalls,&\multicolumn{2}{p{8cm}|}{Adaptive measure: control/Audit/policies}\\
    & Firewalls and VPNs&\multicolumn{2}{p{8cm}|}{Micro-segmentation: smaller and manageable zones}\\
    & Network access control&\multicolumn{2}{p{8cm}|}{Employing AI and ML}\\
         \hline
         \textbf{confidentiality}  & Data protection & Advance data protection& Top data protection\\
       &Prone to data theft/loss & Possible data theft& Hard to data theft\\       
         \hline   
    \end{tabular}
    \label{tab:my_label}
\end{table*}

In \cite{ref14}, the authors developed an intelligent blockchain solution used to maintain secure communication and services in Beyond 5G (B5G) networks. The authors show that moving platform management in B5G is achieved through intelligent connectivity, resource management, and service provisioning. The article highlights the importance of moving networks in NGNs and the necessity to secure such networks in a decentralized fashion. The use of blockchains for device authentication and service provisioning integrity enables autonomous device participation grant and faster authentication of transactions, especially in FANETs. Using private, public, and consortium blockchains within device clusters and between clusters would rapidly speed up the consensus and enables scalability for blockchains. Performance evaluations highlight the significance of malicious service and device detection, such that the use of blockchain-enabled communication and services can decrease malicious activity to near zero.

In \cite{ref15}, a decentralized blockchain-enabled protocol for information sharing is developed for zero trust networks. The solution eases the process of autonomous completion of information sharing in a decentralized manner without relying on trusted third-parties. Fabricated information is identified through smart contracts. Blockchain consensus is maintained using a fair and effective voting mechanism that adopts penalties for misbehaving nodes. Furthermore, temporary keys are distributed for sensitive information encryption. Overall, the developed solution maintains mutual authentication, fairness, autonomy, anonymity, privacy protection, and traceability for zero trust environments. 

\section{Concluding Remarks}
Conventional IDS and IPS solutions have been a favourable security design approach for quite some time. This is mainly due to having devices belong to different network domains. As such, once a device is labeled as trustworthy in the network, any data can be communicated back and forth with the device without deep analysis. The use of passwords and personal identities has been norm for quite some time to grant access to the network. Although the integration of ML solutions to IDS and IPS has provided significant improvements to detect and report malicious activity, reduce risk exposure, and protect data, such centralized conventional systems will not cope with the distribution and decentralization characteristics of NGNs. 

The move towards zero trust will support the distribution and decentralization process of networks and services. Devices may belong to multiple network domains or may not belong to any network domain. As such there needs to be more frequent device validation to ensure a highly controlled network environment. With that, continuous monitoring, auditing, and device behaviour analysis is required. The proposed zero trust architecture have so far shown some shortcomings for NGNs due to their somewhat centralized management nature. For instance, they require centralized key management and access control is managed by a central entity. Such centralized solutions may still pose possible security risks and cannot cope with full levels of scalability.

In 6G, not only devices need to be authenticated and the communicated data needs to be validated, but also the virtual environment such as DTs and the communicated data between DTs in the metaverse will require continuous monitoring. In essence, decentralization is the key towards true zero trust. The solution should provide distributed key management, as well as distributed access control management. Such move towards decentralization will require a number of key enables, most importantly, threat intelligence and the use of intelligent blockchains. Table \ref{tab:my_label} highlights the different features between centralized and decentralized zero trust, as well as conventional IDS and IPS. Lastly, this article highlighted some current trends, potential solutions, and future directions for both threat intelligence and intelligent blockchains for zero trust. We believe that this article will pave the road towards decentralized zero trust for interested researchers.

\balance
\bibliographystyle{IEEEbib}
\bibliography{main}

\begin{IEEEbiographynophoto}
{Ismaeel Al Ridhawi}
is an Associate Professor of Computer Engineering at Kuwait College of Science and Technology and a researcher in the field of wireless communications and networking at the University of Ottawa. 
Contact him at ismaeel.alridhawi@uottawa.ca.
\end{IEEEbiographynophoto}

\begin{IEEEbiographynophoto}
{Safa Otoum}
is an Assistant Professor of Computer Engineering at Zayed University, UAE. She is also a researcher in the field of Network Security and Networking. 
Contact him at safa.otoum@zu.ac.ae.
\end{IEEEbiographynophoto}

\begin{IEEEbiographynophoto}
{Moayad Aloqaily}
is a Research Manager at the Machine Learning Department, Mohamed Bin Zayed University of Artificial Intelligence, Abu Dhabi, UAE. 
Contact him at maloqaily@ieee.org.
\end{IEEEbiographynophoto}

\end{document}